\pdfoutput=1
\documentclass{JINST}
\usepackage{graphicx}
\usepackage{booktabs}
\usepackage{subfig}

\usepackage[stretch=10,final]{microtype}	

\title{Long-Term Testing and Properties of Acrylic for the Daya Bay Antineutrino Detectors}

\author{
	M.\,Krohn$^{a*}$,
	B.\,Littlejohn$^{a*}$,
 	K.\,M.~Heeger$^a$,
 \\
\llap{$^a$}Department of Physics, University of Wisconsin--Madison,\\
  Madison, WI 53706, U.S.A.\\
\llap{$^*$}E-mail: \email{mkrohn@wisc.edu,littlejohn@wisc.edu}}

\abstract{The Daya Bay reactor antineutrino experiment has recently measured the neutrino mixing parameter sin$^22\theta_{13}$ by observing electron antineutrino disappearance over kilometer-scale baselines using six antineutrino detectors at near and far distances from reactor cores at the Daya Bay nuclear power complex. Liquid scintillator contained in transparent target vessels is used to detect electron antineutrinos via the inverse beta-decay reaction. The Daya Bay experiment will operate for about five years yielding a precision measurement of sin$^22\theta_{13}$.  We report on long-term studies of poly(methyl methacrylate) known as acrylic, which is the primary material used in the fabrication of the target vessels for the experiment's antineutrino detectors.  In these studies, acrylic samples are subjected to gaseous and liquid environmental conditions similar to those experienced during construction, transport, and operation of the Daya Bay acrylic target vessels and detectors.  Mechanical and optical stability of the acrylic as well as its interaction with detector liquids is reported.}

\keywords{Radiation and Optical Windows, Gamma Detectors, Detector Design and Construction Technologies and Materials}

\begin{document}

\tableofcontents
\clearpage

\section{Introduction}
The Daya Bay reactor antineutrino experiment is located at the Daya Bay nuclear power complex near Shenzhen, China. The experiment presently consists of six cylindrical antineutrino detectors (ADs) located in three underground experimental halls at distances ranging from 300--2000\,m from six nuclear reactor core~\cite{DYBTDR}.  Installation of two additional detectors is planned for fall 2012. The detector locations are chosen to optimize the sensitivity of the experiment to neutrino oscillations and to allow the measurement of the neutrino mixing angle $\theta_{13}$. The Daya Bay experiment detects antineutrinos through the inverse beta decay reaction $\overline{\nu}_{e}+p \rightarrow e^+ + n$. Observation of a deficit of observed electron antineutrinos in the far detectors is a signature of neutrino oscillation and allows the measurement of sin$^22\theta_{13}$~\cite{DYBPRL}.

\begin{figure}\centering
       \includegraphics[height=3.5in]{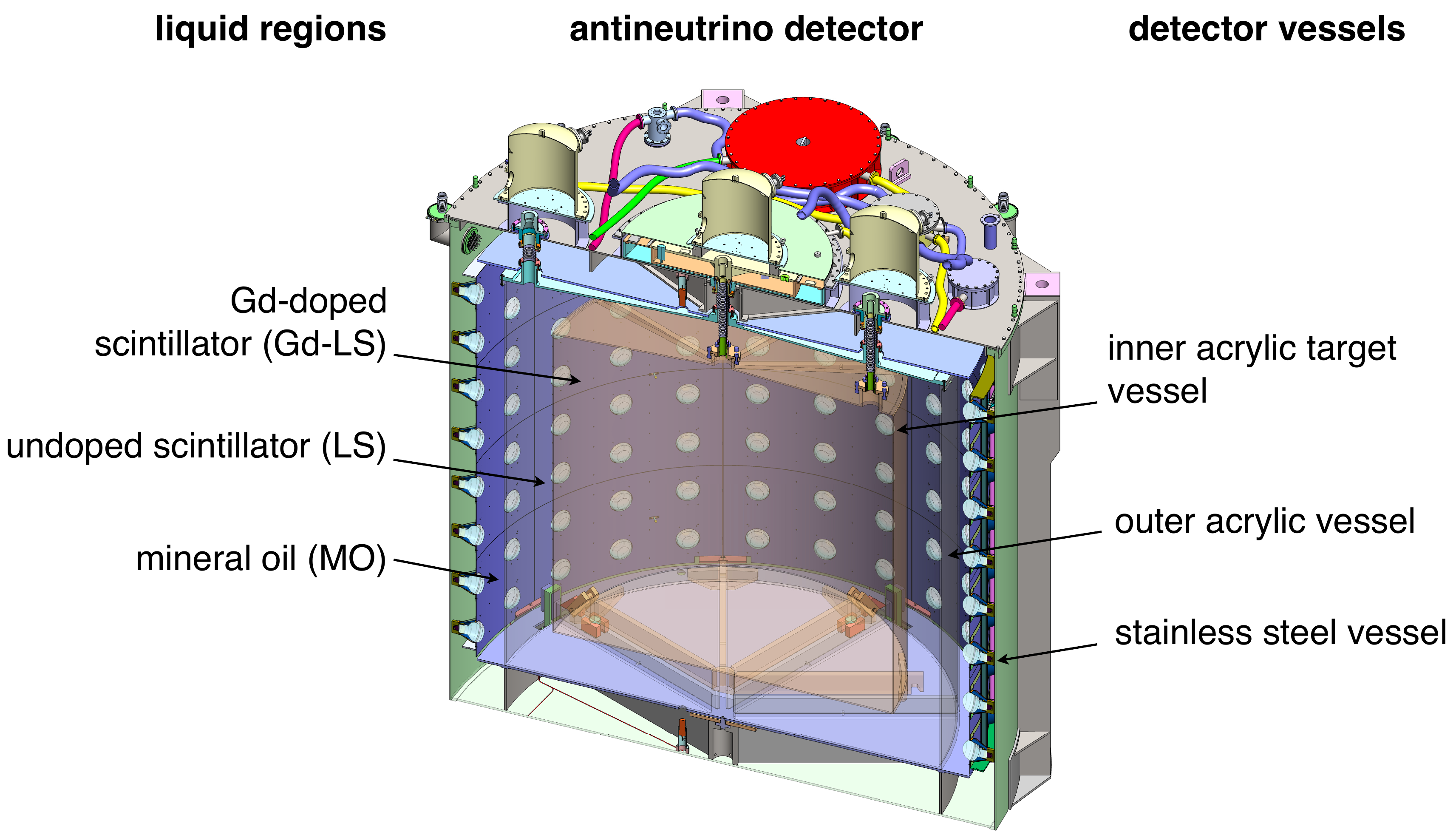}
        \caption{Cutaway view of an antineutrino detector, with the inner acrylic vessel containing the Gd-LS, the outer acrylic vessel containing the undoped LS, and the stainless steel vessel containing the mineral oil region and PMTs.}
        \label{fig:AD}
\end{figure}

The Daya Bay antineutrino detectors, shown in Figure~\ref{fig:AD}, consist of a nested arrangement of upright cylinders forming 3 concentric zones. The detectors have an outer diameter and height of 5\,meters. The detector target consists of 20 tons of gadolinium-doped liquid scintillator (Gd-LS) and is contained in a 3-m diameter inner acrylic vessel (IAV). This inner vessel is surrounded by a 4-m diameter outer acrylic vessel (OAV) holding 21 tons of undoped liquid scintillator acting as a gamma catcher for events with with high-energy gammas leaking out of the inner target.  The design and construction of the Daya Bay acrylic vessels (AVs) is described in detail in \cite{BryceAV}.  The liquid scintillator is based on the organic solvent linear alkyl benzene (LAB), and is described in more detail in \cite{GdLS}.  The region beyond the OAV is filled with a mineral oil buffer, instrumented with 192 photomultiplier tubes (PMTs) and bounded by a stainless steel vessel (SSV).  The PMTs detect scintillation light from the inner regions, using the total detected light to determine the energy of particle interactions in the detector.

To ensure the long-term stability of the experiment, long-term ($>5$~year) compatibility between the scintillating liquids and their containing vessels is a basic technical requirement of the detector design.  No long-term degradation or instability of detector liquid samples has been found after several years' exposure to acrylic~\cite{Minfang}.  Conversely, this paper investigates the long-term impact on the acrylic from constant exposure to the detector fluids over several years.  In particular, we investigate the acrylic's ability to absorb detector liquids, which can affect the total target mass of the detector, a fundamental input to measurement of sin$^22\theta_{13}$.  Subsequent changes in mechanical and optical properties resulting from this absorption are also examined.

In addition, we investigate the impact of the changing environmental conditions during the fabrication, transport, and installation of the acrylic vessels on the properties and long-term stability of the acrylic. Many stages in the AV production and delivery cycle may potentially introduce changes in the optical, mechanical, and material properties of the Daya Bay acrylic. Possible changes from UV light exposure during production, transport, and assembly were documented in~\cite{BryceDeg}.  Differences in geometric, optical, and radioactivity arising from differences in production and assembly were given in \cite{BryceAV}. This paper seeks to identify any possible changes in properties arising from shipping, storage, and deployment of the AVs in the operational detectors, important aspects not discussed in previous papers.

In particular, we investigate the mechanical, optical, and moisture-content effects of the following acrylic environments:
\begin{itemize}{}
\item Large temperature fluctuations experienced during vessel transport.
\item Exposure of the vessels to dry nitrogen environments during storage after assembly and before filling with detector fluids.
\item Exposure of the vessels to target liquids during the multi-year lifetime of the experiment.
\end{itemize}

Understanding these effects will be useful not only to the Daya Bay experiment, but also to future neutrino and dark matter experiments utilizing acrylic as a detector material.

\section{Methods of Testing and Characterization}

\subsection{Acrylic Sample Preparation}

Acrylic properties were investigated by submitting samples of production acrylic to a variety of tests. The sample acrylic was taken from production slush-cast acrylic sheets produced by Reynolds Polymer Technology, Inc.~\cite{Reynolds}, and then cut into smaller pieces of equal size, $8~$in$~\times~1.5~$in$~\times~0.375~$in.  After cutting, sample surfaces were polished to 1200~grit with abrasive paper, and then annealed according to~\cite{Stachiw}, removing any residual stress introduced during sample preparation.

\subsection{Optical Stability and Transmittance}

\begin{figure}[t]\centering
        \includegraphics[height=2.5in]{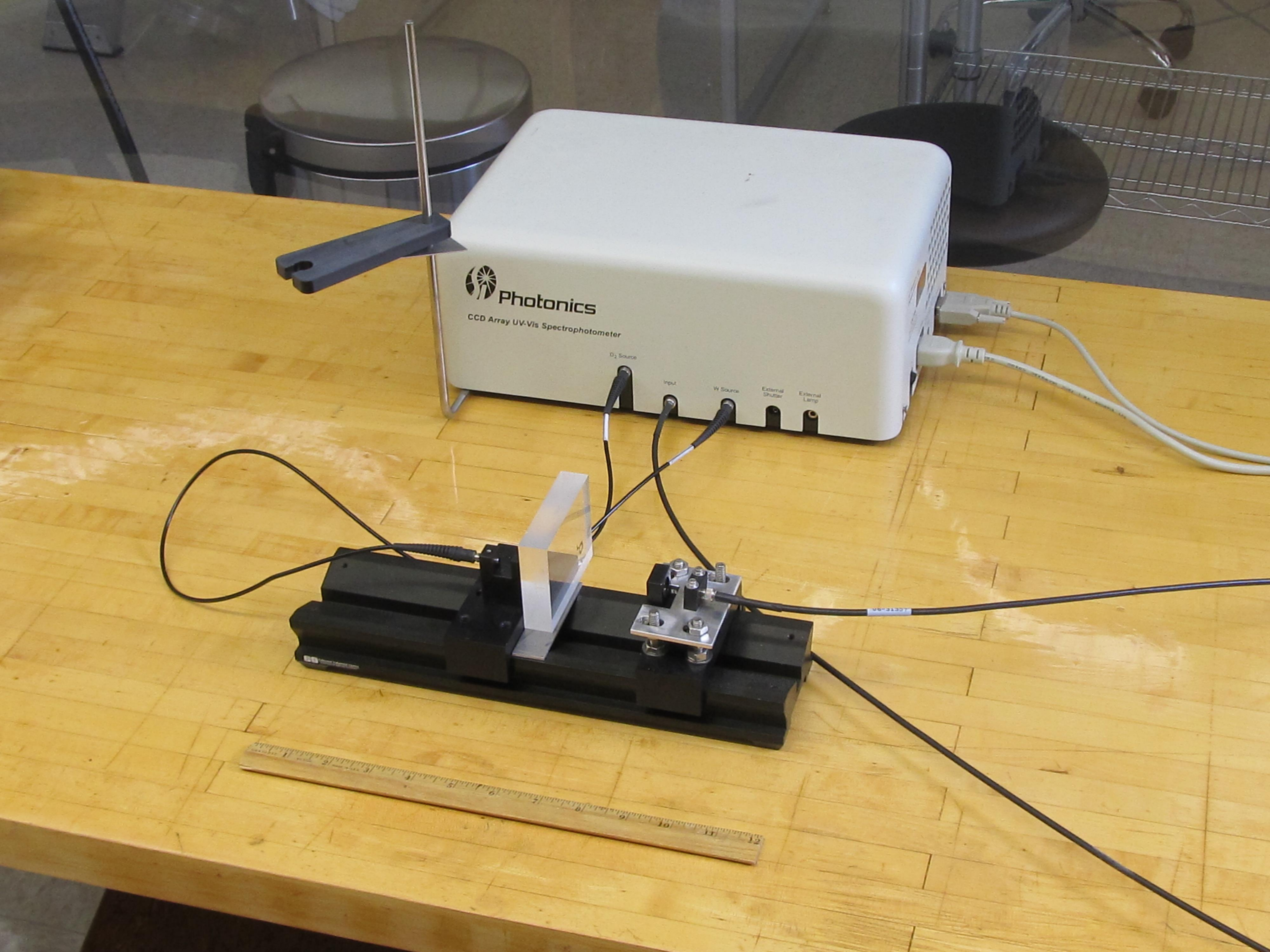}
        \caption{A photograph of the transmittance measurement setup.  UV and visible light generated by tungsten and deuterium lamps in the spectrometer housing (back) are fed through a fiber-optic cable to the black sample holder (front).  After exiting the fiber and traveling through the sample, the light beam impinges upon another fiber-optic cable on the opposite side of the sample.  Collimating lenses reduce the dispersion of the light beam.  This output light is fed into a CCD array in the spectrometer housing to analyze the wavelengths present after sample traversal.}
        \label{fig:Transmittance}
\end{figure}

The optical stability of various samples was tested using an SI Photonics 400 Series UV-Vis spectrometer~\cite{SIPhotonics}, which measured the percent transmittance of various light wavelengths through each sample.  The transmittance measurement setup is shown in Figure~\ref{fig:Transmittance}.  Transmittance was measured before and after exposure of each sample to a particular environment and then compared to determine that environment's effect on the optical properties of the acrylic.  Samples were labelled to achieve consistent sample orientation from scan to scan.  The systematic uncertainty of the UV-Vis spectrometer was determined through repeated testing of similar samples to be 2--3\%; this uncertainty arises from slight deviations in sample placement and orientation between different transmittance scans, and from minor variations in sample surface flatness.  

\subsection{Mechanical Stability and Stress Testing}
\label{subsec:stress}

\begin{figure}[t]
\subfloat[Top-down schematic view of the stress-testing setup.]{\label{subfig:StressSetup}%
  \includegraphics[trim=5cm 4cm 6cm 4.4cm, clip=true, height=0.37\textwidth]{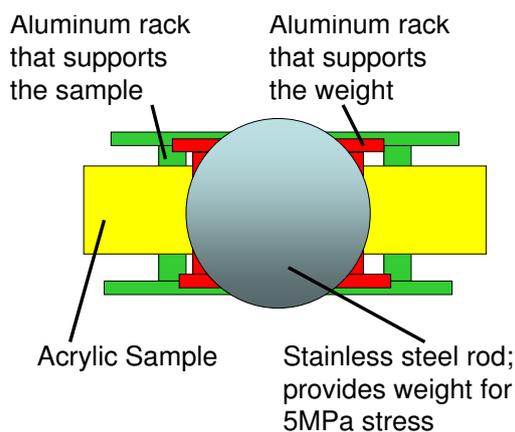}%
}\hfill
\subfloat[Photograph of the stress-testing setup.]{\label{subfig:StressSample}%
  \includegraphics[height=0.37\textwidth]{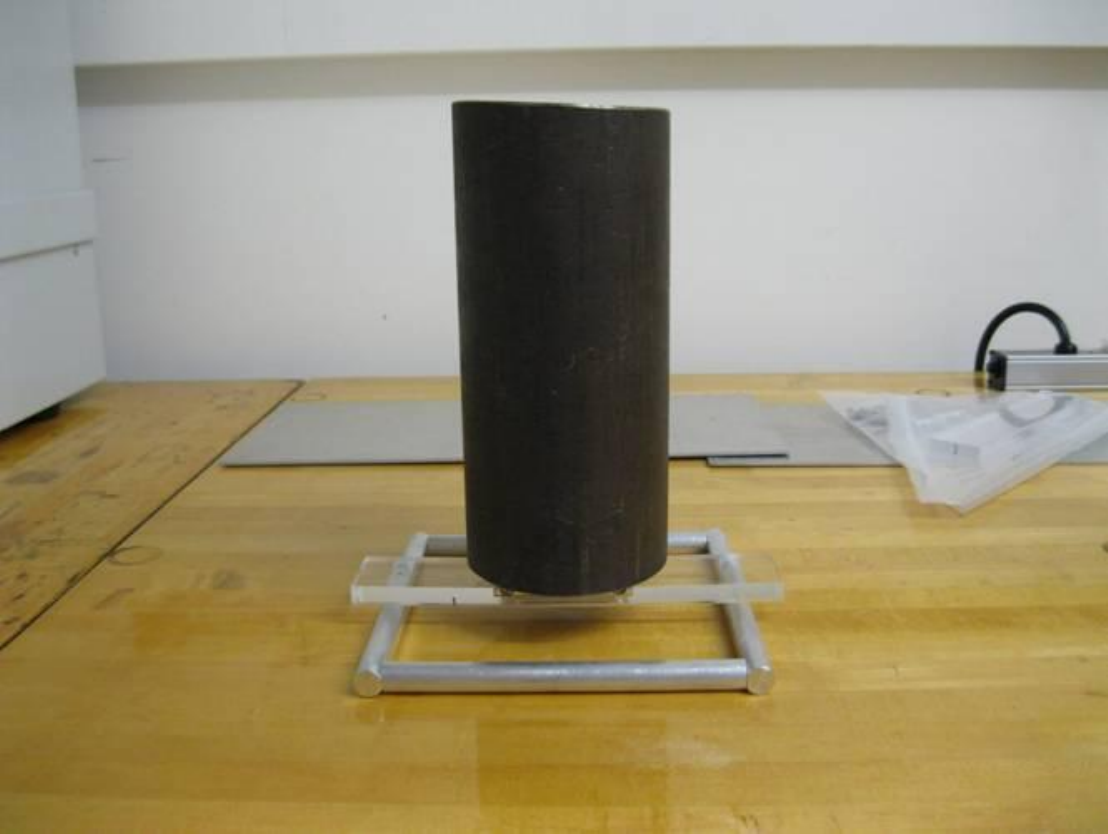}%
}%
\caption{Stress test setup following ASTM D6272. (a) Schematic top view.  The acrylic sample (yellow) rests between two Aluminum frames shown in green and red. A cylindrical weight is applied to the top of the red frame which exerts a bending force and resulting stress on the acrylic sample. The bending geometry is defined by the size and spacing of the aluminum frames. (b) Photograph of the test setup.}
\label{fig:StressSetup}
\end{figure}

The mechanical stability of the acrylic samples was measured by stressing the samples according to the ASTM D6272 standard \cite{ASTM}. The set-up of this stressing procedure, shown in Figure~\ref{fig:StressSetup}, depends on a standardized geometry to achieve a desired level of stress in the acrylic sample.  Acrylic samples rest between a large bottom frame and a smaller top frame.  A steel weight is then placed on the top frame, providing the desired level of stress.  The total stress experienced by the acrylic is given by
\begin{equation}
S = PL/bd^2,
\end{equation}
where $S$ is the stress applied to the sample in MPa, $P$ is the value of the weight in N, $L$ is the distance between the two rods of the rack the acrylic samples rests on in mm, $b$ is the width of the acrylic sample in mm, and $d$ is the depth of the acrylic sample in mm. For our experiment, P~=~113.6~N, L~=~152.4~mm, b~=~38.1~mm, and d~=~9.5~mm, which produces 5.0 MPa of stress in the acrylic sample.  This stress value corresponds to the 10-year maximum stress limit guideline followed by Reynolds for all commercial operations and tested in~\cite{BryceAV}.  If exposing stressed samples to a given environment alters appreciably the acrylic's mechanical characteristics, the samples may exhibit signs of cracking or crazing over time.

\subsection{Sample Mass and Dimensions}

The liquid absorption and moisture content in the acrylic samples was characterized by measuring the mass and dimensions of the samples.  Absorption of liquids into the acrylic vessels can affect both the mechanical properties of the acrylic as well as the determination of the target mass inside the ADs.  Absorption into the acrylic vessel material would result in an effective decrease of the liquid level in the detectors' overflow tanks, which is used to determine the target mass during data taking.

Micrometers with an accuracy of 0.0001~in were used to measure the length, width, and depth of each samples.  A Sartorius scale~\cite{Sartorius} with an accuracy of 0.1~mg was used to measure each sample's mass.  For samples submerged in detector liquids, additional mass uncertainty was introduced by small variations in the sample mass measurement preparation process.  This additional uncertainty will be discussed in more detail in Section 3.3.

Using the original mass $m$ and dimensions $l$, $w$, $d$ of each sample, the fractional changes  ${\Delta m}/{m}$, ${\Delta l}/{l}$, ${\Delta w}/{w}$, and ${\Delta d}/{d}$ from exposure to a given liquid or gas environment were determined. These properties were measured in a variety of tests that simulated the multiple environments the AVs experienced during shipping and operation.  Any changes in these quantities was indicative of changes in the concentration of absorbed liquids and moisture in the samples.

\section{Acrylic Tests}

\subsection{Temperature Cycling}

\begin{figure}[tb]\centering
        \includegraphics[height=3in]{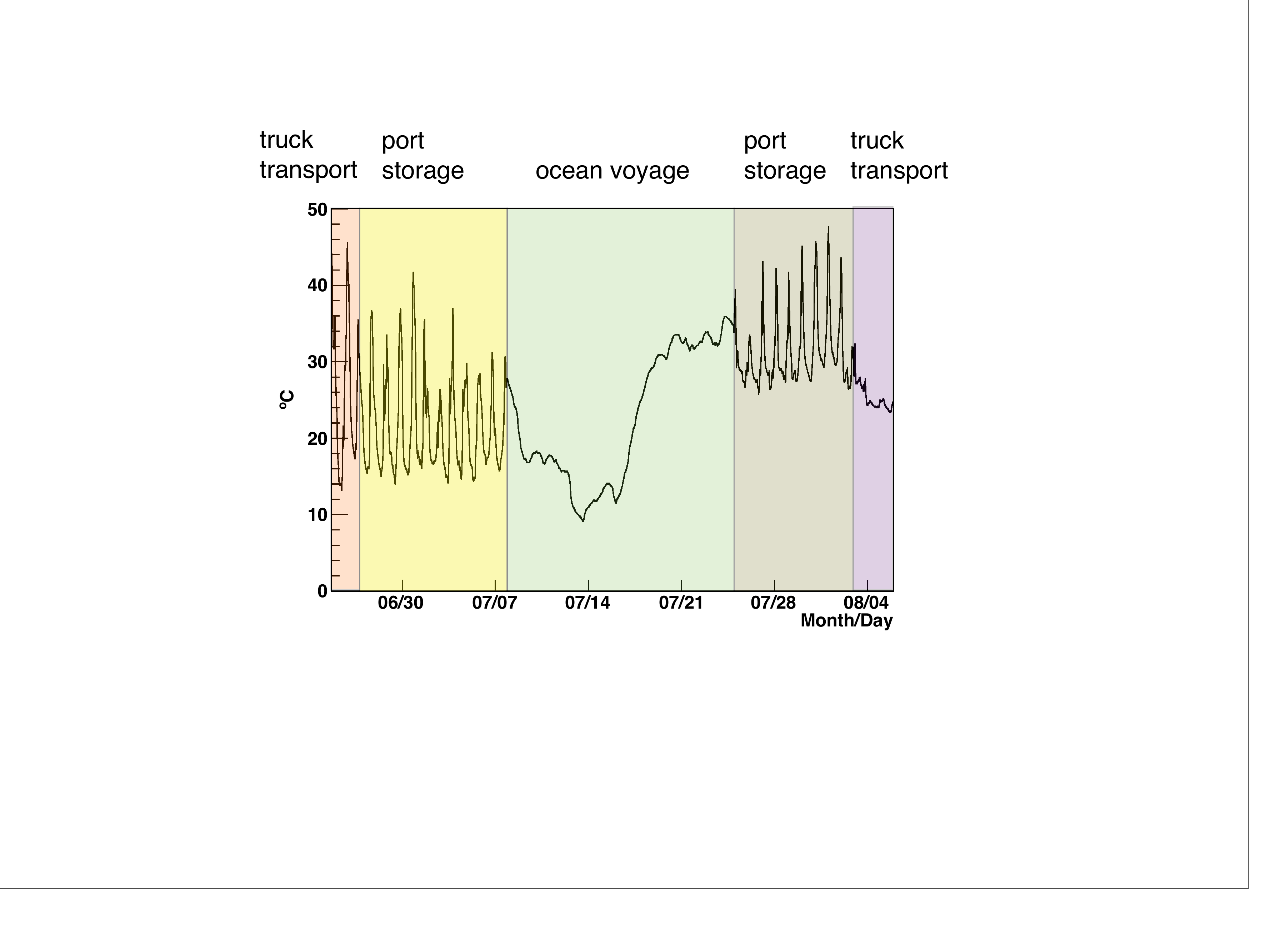}
        \caption{Typical temperature measurements experienced during transport of the acrylic vessels from the US to China in 2010. The colored regions represent five periods: transport by truck in the US (red), during storage at port in Long Beach, CA (yellow), during sea transport (green), at port in Yantian, China (gray), and en route to Daya Bay and in the Surface Assembly Building (purple). In our laboratory tests we simulated the temperature cycles experienced during the period from about 7/18 to 8/04. Figure adapted from \cite{BryceAV}.}
        \label{fig:TemperatureCycling}
\end{figure}

During transport and shipping the acrylic vessels were subject to temperature changes due to environmental and diurnal variations. The minimum and maximum temperature variations, as measured by on-board temperature sensors at various positions on the vessel shipping frame, ranged from 10 to 46 $^\circ$C with typical temperatures in the range of 14 to 35 ${^\circ}$C\null. Figure~\ref{fig:TemperatureCycling} illustrates the temperature variations during the shipment of the outer acrylic vessels for the first two Daya Bay detectors. In order to simulate these temperature changes experienced by the AVs, two acrylic samples were placed in a programmable convection oven. The temperature of the oven was cycled to mimic the measured temperature during transportation. While undergoing these temperature changes the samples were also stressed using the testing setup described in~Section~\ref{subsec:stress}.  After finishing the cycling regimen, the samples were inspected for micro-cracks and tested for changes in transmittance.  These results were compared to a sample that was stressed in the absence of temperature-cycling to observe any difference in properties.

\begin{figure}[htb]\centering
        \includegraphics[height=3in]{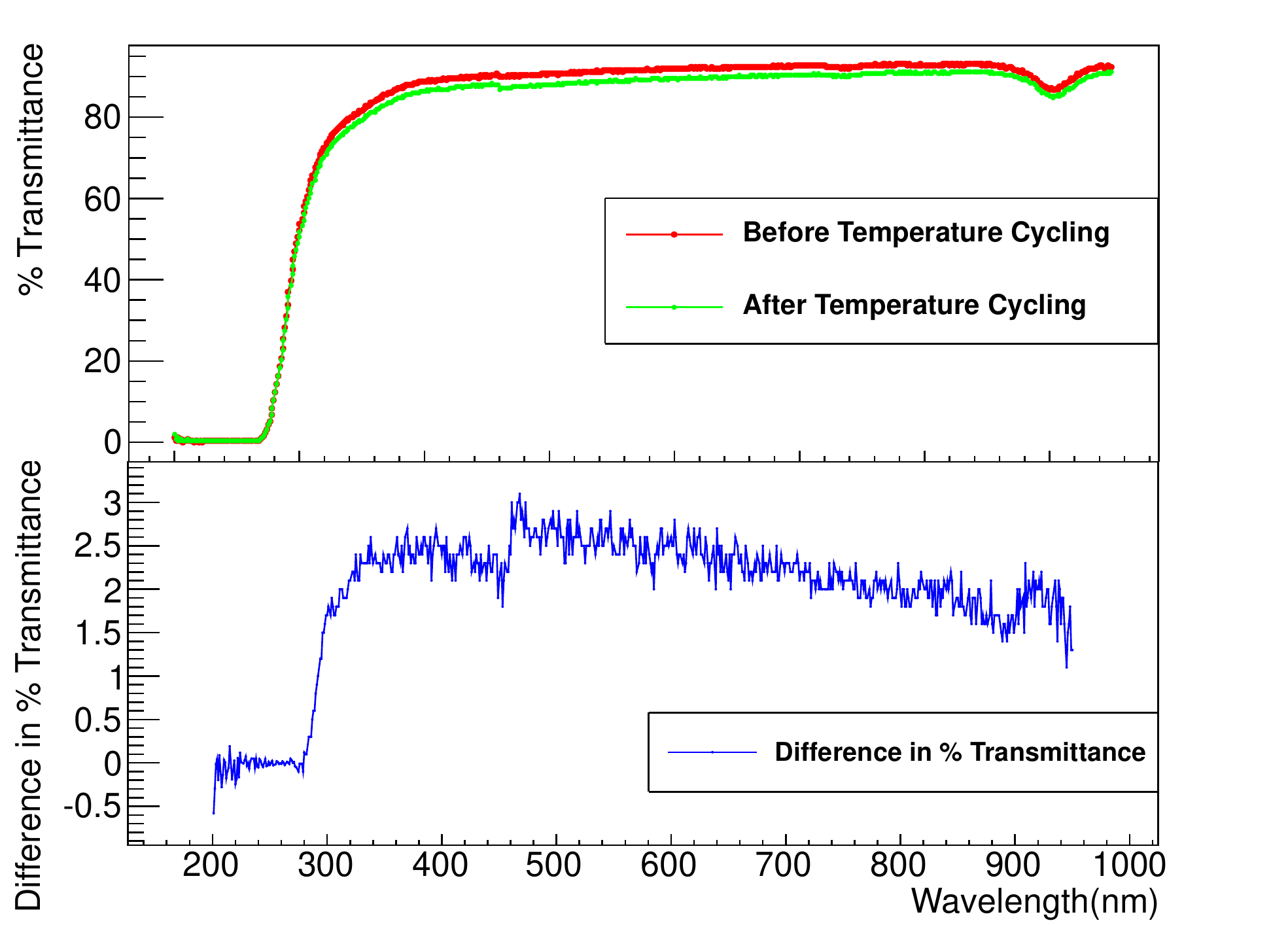}
        \caption{Comparison of acrylic sample transmittance before and after temperature cycling.  Overall differences between initial and final spectra are smaller than the systematic uncertainty of the spectrometer, indicating no significant change in optical properties from temperature cycling.  The measured step in transmittance at 460~nm arises from the switchover between the tungsten and deuterium lamps in the spectrometer.}
        \label{fig:CycledTransmittance}
\end{figure}

We found that temperature cycling did not cause any detectable cracking or crazing of the samples, indicating no significant change in acrylic mechanical properties. As shown in Figure~\ref{fig:CycledTransmittance},  the transmission spectrum before and after cycling was unchanged within the spectrometer measurement uncertainty. These results show that the AV transport procedure from the US to China introduces no significant changes in acrylic properties.

\subsection{Exposure to Dry Air Environments}

The first two antineutrino detectors' acrylic vessels were stored in a nitrogen environment derived from liquid nitrogen boil-off for several weeks before they were filled with the detector liquids, possibly leading to unquantified changes in the acrylic's moisture content as well as mechanical and optical properties.  Furthermore, acrylic vessels for the six subsequent detectors were filled shortly after completion of detector assembly, resulting in a possible relative difference in AV moisture and mechanical and optical properties between detectors.  In order to investigate the nature and magnitude of this difference, two separate dry-environment acrylic tests were conducted.

\subsubsection{Oven-Drying of Acrylic}
\label{subsec:DryEnv}

\begin{figure}\centering
        \includegraphics[height=2.5in]{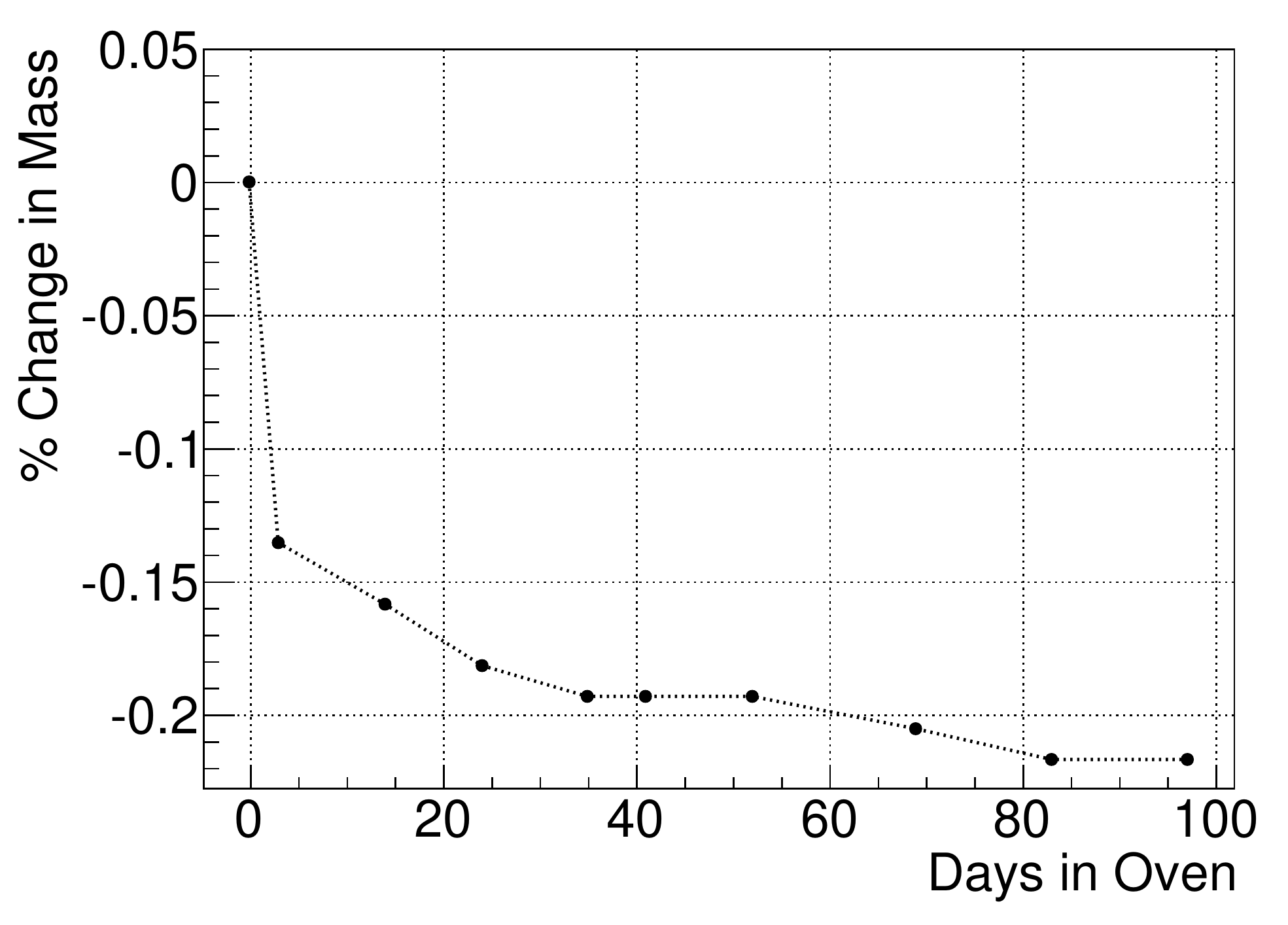}
        \caption{Percent change in mass of an acrylic sample consistently baked at 60$^{\circ}$C. The initial large drop in mass is likely caused by water evaporation, while the slow subsequent decrease in mass might be the result of degradation of the acrylic chemical structure by extreme dryness or elevated temperatures.  The initial total water content of the sample appears to be 0.1--0.2\% by mass.}
        \label{fig:BakedWeight}
\end{figure}

The first test simulated a dry environment by placing acrylic samples in a programmable convection oven operating at $60\,^{\circ}\mathrm{C}$, well below acrylic annealing temperatures.  While this does not replicate the environment experienced by the stored AVs, the oven-drying results in complete dry-out of the acrylic, and thus provides an extreme bounding case on possible changes in acrylic properties from exposure to dry environments.  Acrylic samples were stressed and baked for a four-week period, and were periodically removed to test their mass, dimensions, and transmittance.  The change in sample mass as a function of time can be seen in Figure~\ref{fig:BakedWeight}.

This data shows a quick initial exponential loss in mass, likely from evaporation of water in the acrylic, followed by a longer-timescale loss, possibly from slow degradation and evaporation of the acrylic chemical structure from either extreme dryness or elevated temperature.  From this test, one can determine that the equilibrium water content of acrylic in average room air humidity is 0.1--0.2\% by mass.

\begin{figure}\centering
        \includegraphics[height=3in]{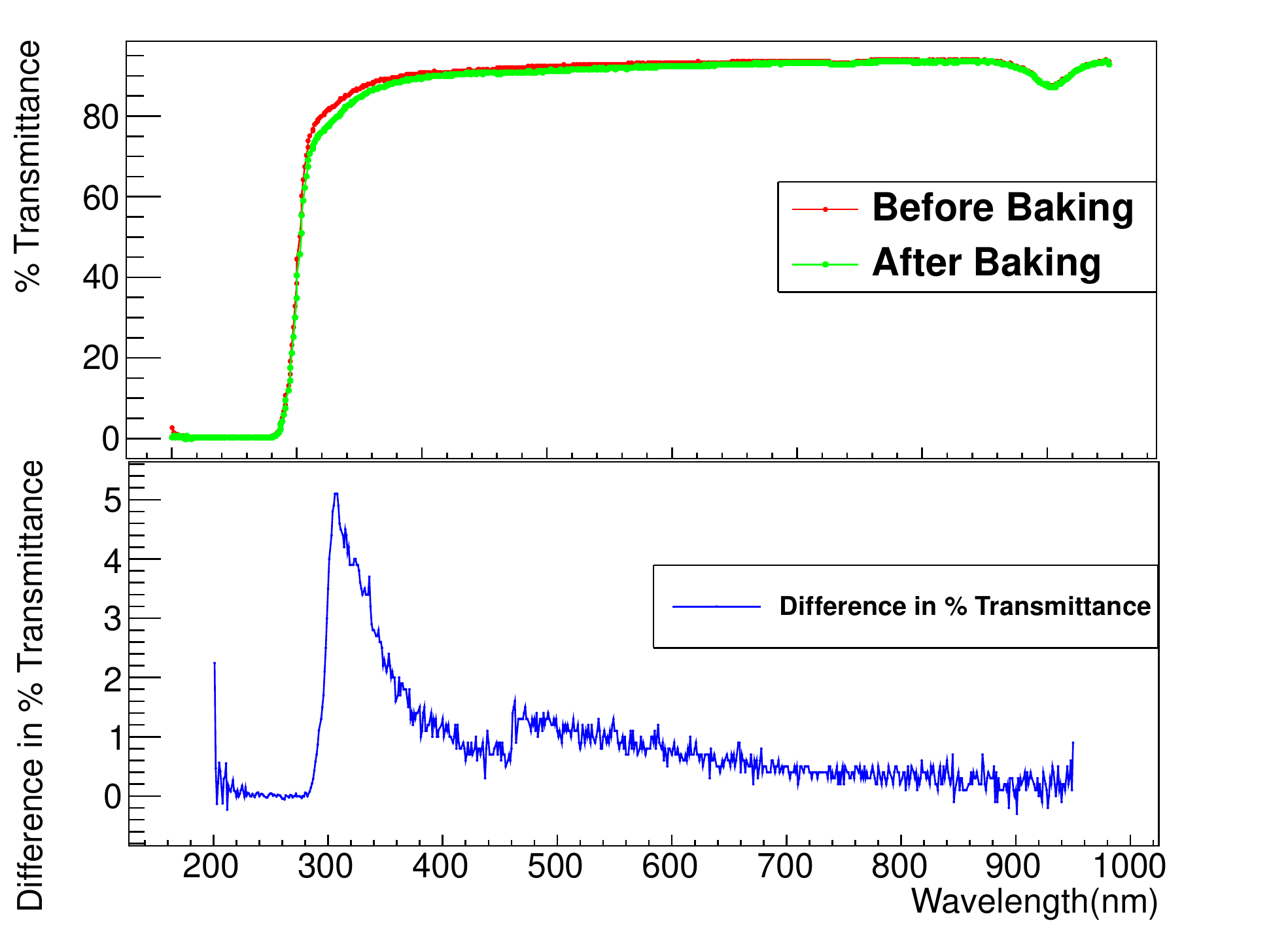}
        \caption{Transmittance of an acrylic sample before and after four weeks of consistent baking at 60$^{\circ}$C.  A small change in transmittance is visible below 400~nm, below the range of light production and transmittance for the Daya Bay liquids.  The measured step in transmittance at 460~nm comes from the switchover between the tungsten and deuterium lamps in the spectrometer.}
        \label{fig:BakedTrans}
\end{figure}

Despite this change in mass, no cracking or crazing of the baked samples was visible at the end of the test, indicating no appreciable change in mechanical properties.  The change in transmittance before and after baking is shown in Figure~\ref{fig:BakedTrans}.  Significant but small changes in transmittance were visible only below 400~nm, below the range of light production and transmittance for the Daya Bay liquids.  Thus, from this baking test, we conclude that severely dry acrylic vessels exhibit no substantial change in optical or mechanical properties, and reduce in mass by 0.1--0.2\%.  This corresponds to a loss of at most 2~kg for the IAV and 4~kg for the OAV compared to vessels of normal moisture level.  In comparison, the as-measured mass of the inner and outer acrylic vessels for AD1 at the time of installation into the AD were 907~kg and 1847~kg, respectively.

After this four-week baking period and subsequent measurements, dry samples were placed in various detector liquids and monitored.  The results of these subsequent tests are described in Section~\ref{subsec:Dry}.

\subsubsection{Drying of Acrylic in a Nitrogen Gas Environment}

\begin{figure}\centering
        \includegraphics[height=3in]{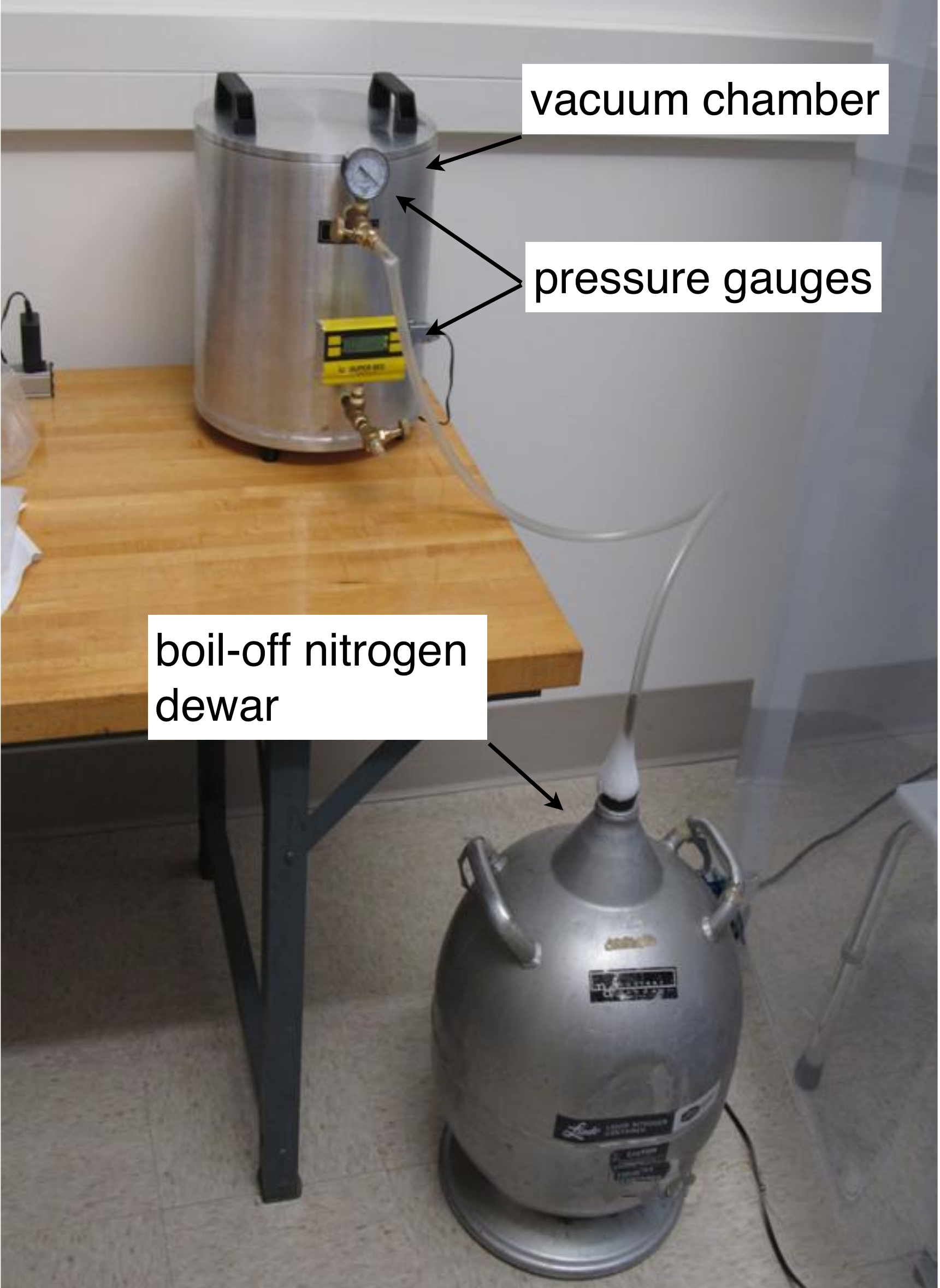}
        \caption{Liquid nitrogen boil-off environment test setup. A vacuum chamber connected to the boil-off from a liquid nitrogen dewar is used to simulate the long-term effects nitrogen cover gas on the acrylic vessels.}
        \label{fig:NitrogenExperiment}
\end{figure}

The second dry-environment test attempted to more closely simulate the nitrogen storage environment by placing acrylic samples in a nitrogen-flushed Abbess vacuum chamber~\cite{Abbess}, as shown in~Figure~\ref{fig:NitrogenExperiment}.  The nitrogen environment was achieved by connecting a liquid nitrogen dewar to the input of the vacuum chamber and allowing the boil-off to flow through the chamber.  As with the stored AVs, in this environment moisture could constantly evaporate from the sample and be carried away in the continuously flushed nitrogen gas.  These samples were also stressed according to ASTM D6272 and visually checked through a clear acrylic portal on the top of the chamber for crazing or cracking.  Samples were removed from the chamber periodically to check the mass and dimensions of the samples.  When the sample was replaced in the chamber after measurement, the dry nitrogen environment was quickly restored by conducting a thorough flush with bottled nitrogen gas.

\begin{figure}\centering
        \includegraphics[height=2.5in]{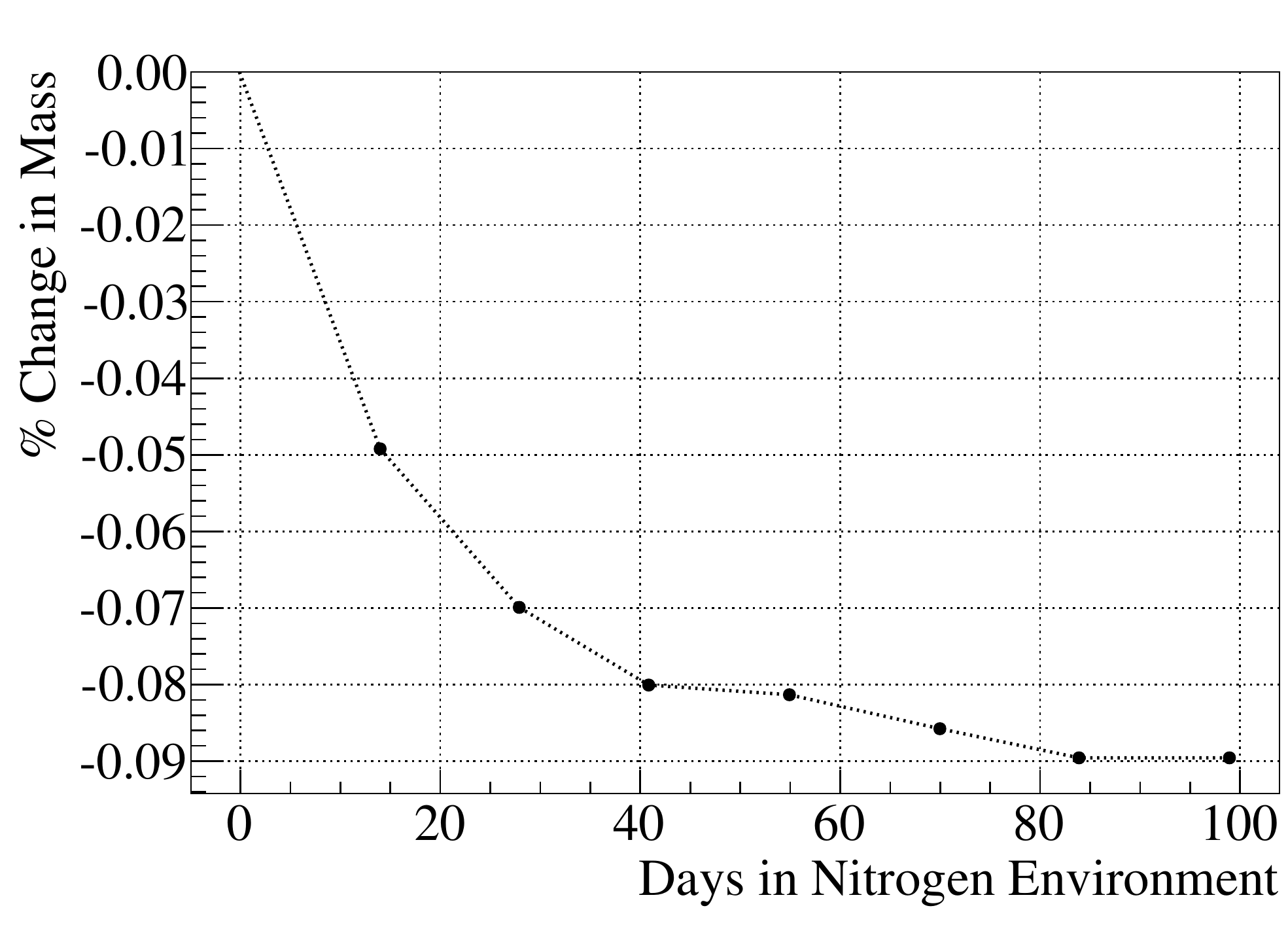}
        \caption{The percent change in mass of an acrylic sample in a nitrogen boil-off environment.  Error bars are smaller than the plotted points.}
	\label{fig:NitrogenMass}
\end{figure}

It was found that the nitrogen environment did not cause any cracking or crazing in the sample, indicating no significant change in acrylic mechanical properties.  In addition, no change in optical properties larger than the spectrometer systematic error were observed.  As shown in Figure~\ref{fig:NitrogenMass}, the sample decreased in mass by $0.0897\pm 0.0002\%$ over the course of 100 days.  This result indicates a total equilibrium water weight of acrylic in average room air humidity of around 0.1\%, in reasonable agreement with that seen in the oven-dried sample.

Given a comparable rate of nitrogen flow through the stored acrylic vessels, it is reasonable to expect a similar rate of water evaporation in the AVs.  Thus, storage of AVs in a dry nitrogen environment on month-long time scales will significantly deplete the AVs of moisture, but will not impact the AV optical or mechanical properties.  Further tests presented in the following section will demonstrate how this dryness impacts an AV's ability to absorb detector liquids.

\subsection{Interaction with Detector Liquids}

The interaction of acrylic with the detector liquids was also investigated.  Because of different storage conditions in varying degrees of dry atmosphere (air or dry nitrogen), the production acrylic vessels for the Daya Bay detectors likely have different moisture levels.  For this reason, acrylics of various moisture levels were tested.

\subsubsection{Acrylic in Average Room Air Humidity}
\begin{figure}[t]\centering
        \includegraphics[trim=0cm 6.5cm 0cm 3.5cm, clip=true, height=2.5in]{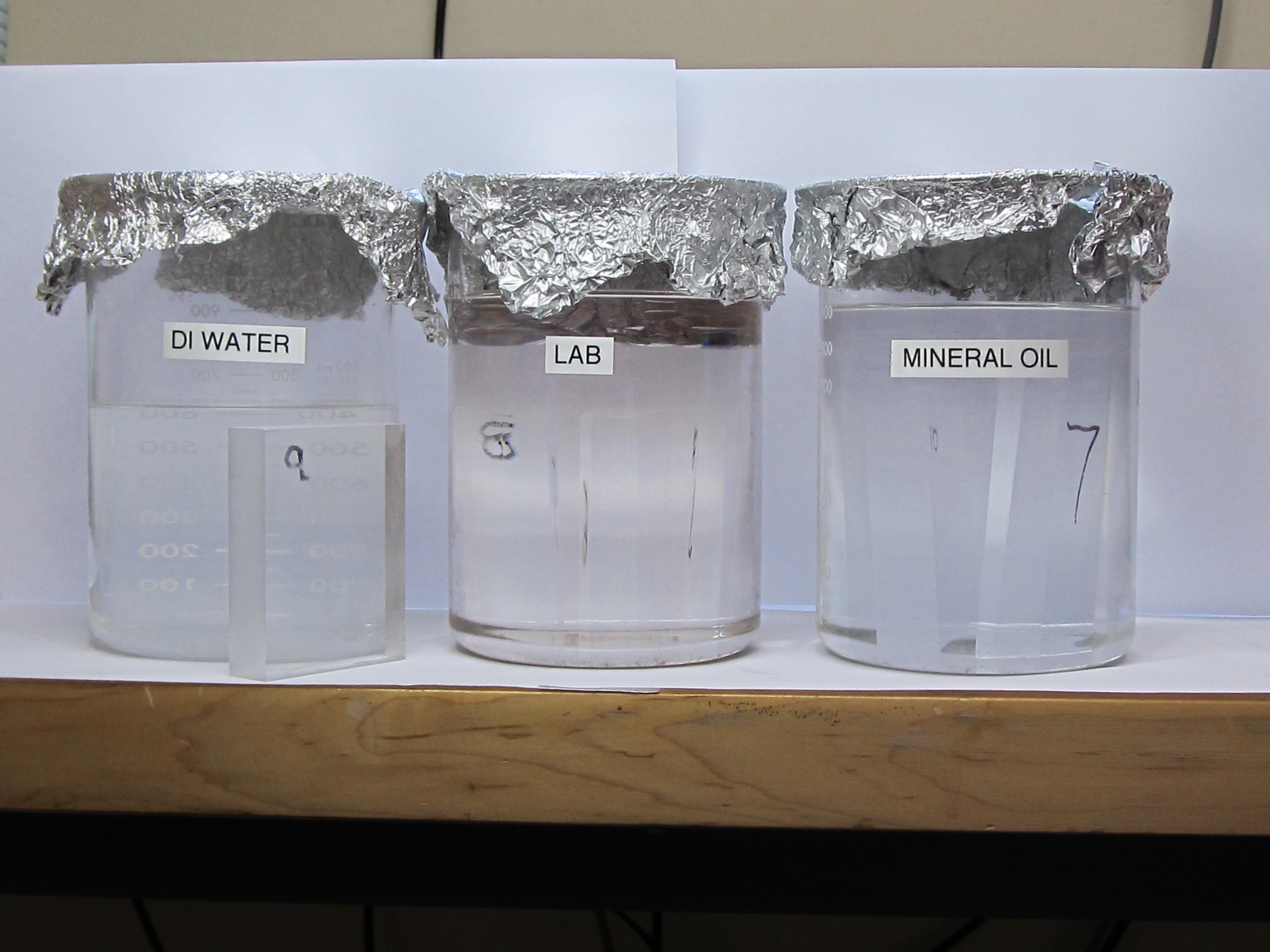}
        \caption{Liquid immersion and soak test setup of acrylic samples in the detector liquids LAB and mineral oil, as well as in deionized water.}
        \label{fig:Liquids}
\end{figure}

A first test was conducted by placing three acrylic samples exposed to a typical laboratory environment of roughly 50\% humidity and room temperature into each of the detector liquids LAB and mineral oil, as shown in Figure~\ref{fig:Liquids}.  Control samples were soaked in deionized (DI) water because acrylic is known to absorb roughly 2\% of its mass in DI water at full saturation, according to R\&D tests done for the SNO experiment~\cite{SNO}.

Every two weeks the mass, dimensions, and transmittance were measured to test for any changes in acrylic properties.  Samples were prepared for measurement with a standardized procedure: samples were removed from the liquid, quickly washed with surface cleaner and water, and then dried using delicate tissue wipes. Each sample was weighed, measured, and placed back in the liquid within 10 minutes of taking it out of the liquid.  The variation in the time taken during each preparation step can introduce biases into the measurement of each sample's mass, as samples could change their moisture content through absorption or evaporation in the room air environment where tests were conducted.  To quantify this uncertainty, the mass of a few acrylic samples in MO and LAB were measured daily while varying the washing, rinsing, and drying time before measurement.  As the length of each step in the preparation cycle was varied by up to a minute, the sample's measured mass varied by up to 0.005~g. This corresponds to an additional 0.01\% uncertainty in each mass measurement.

\begin{figure}[htb]\centering
\subfloat[Change in mass of submerged samples.]{\label{subfig:MassChange}%
  \includegraphics[trim=0cm 0cm 0.03cm 1cm, clip=true, width=0.48\textwidth]{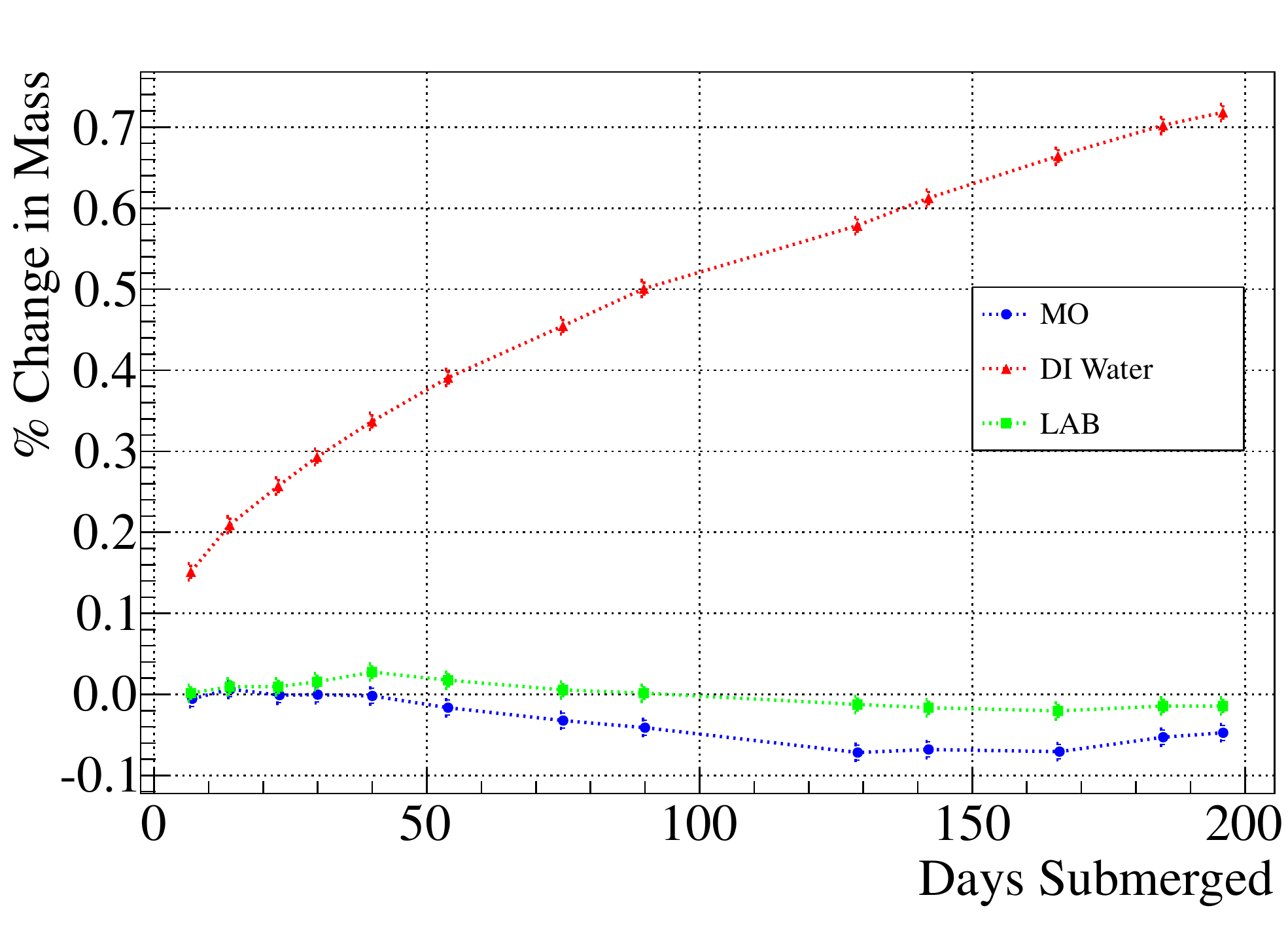}%
}\hfill
\subfloat[Change in length of submerged samples.]{\label{subfig:LengthChange}%
  \includegraphics[trim=0cm 0cm 0cm 1cm, clip=true, width=0.48\textwidth]{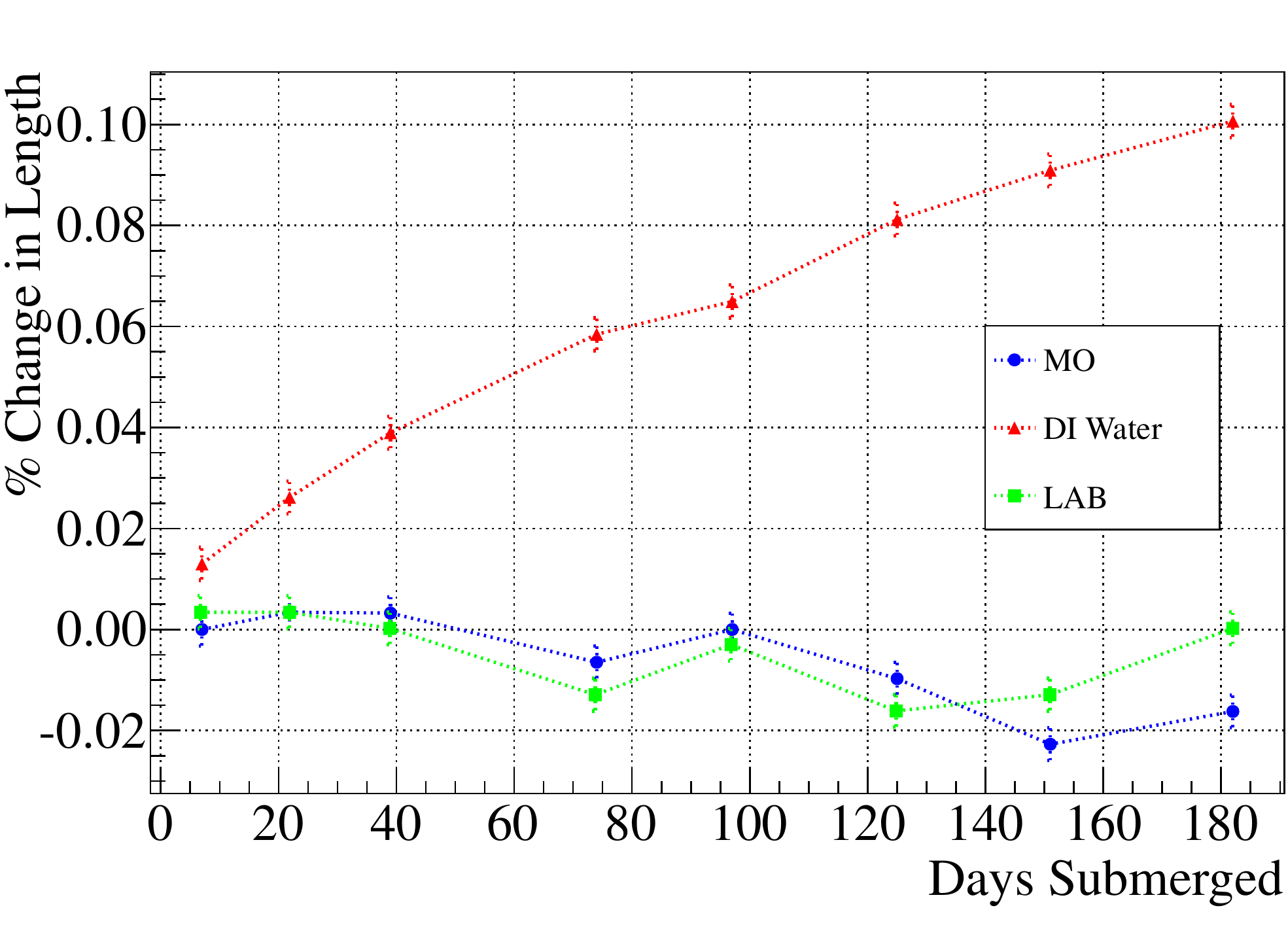}%
}\\
\subfloat[Change in width of submerged samples.]{\label{subfig:WidthChange}%
  \includegraphics[trim=0cm 0cm 0cm 1cm, clip=true, width=0.48\textwidth]{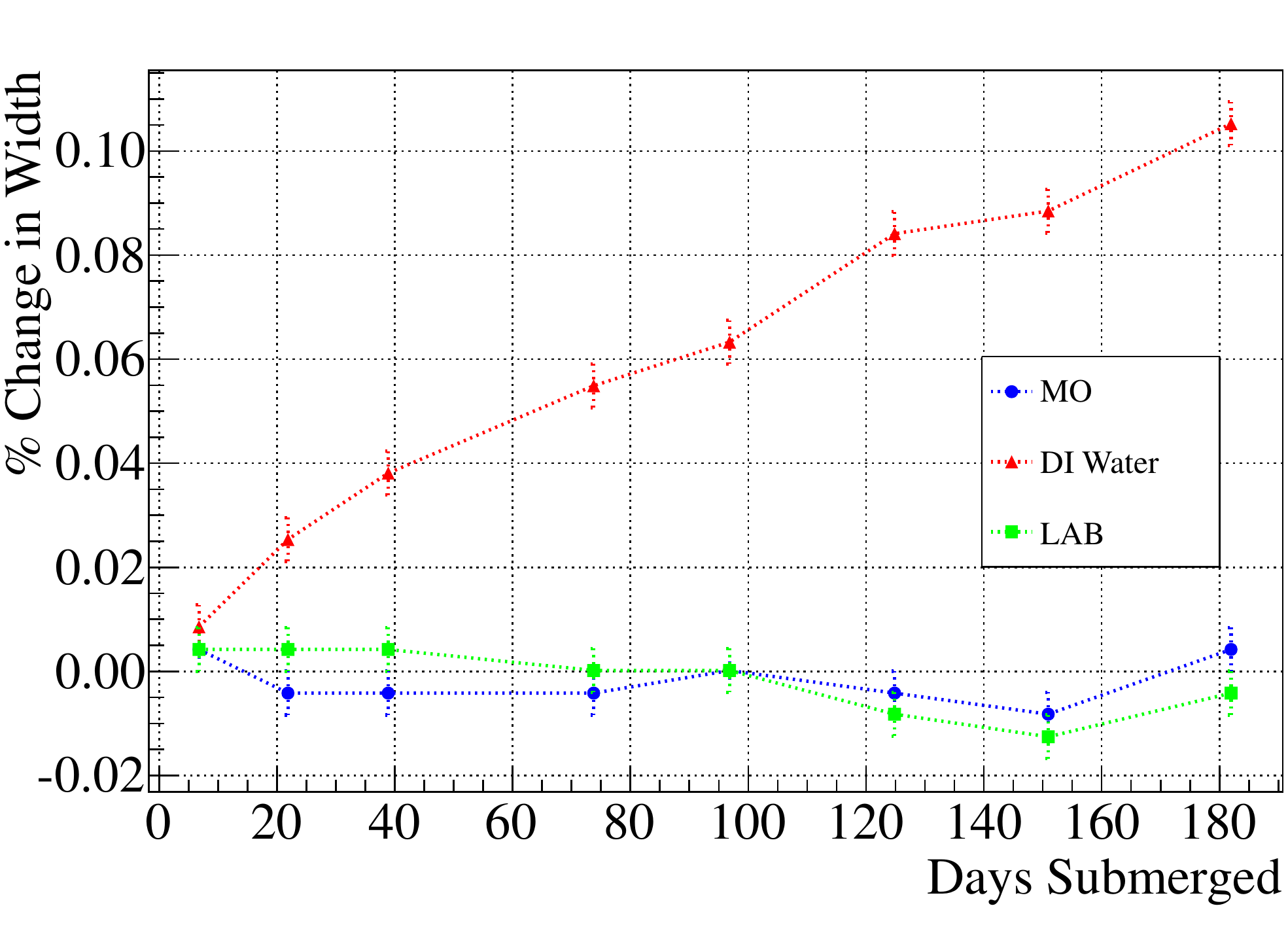}%
}%
\caption{The percent change in the mass (top left), length (top right), and width (bottom) of room-air moisture level acrylic samples submerged in DI water, LAB, and mineral oil.  Water-submerged samples show a clear increase in mass and lengths, while samples submerged in the Daya Bay detector liquids MO and LAB do not.}
\label{fig:AirSamples}
\end{figure}

Individually for every liquid, each of the three submerged room-air moisture level acrylic samples showed similar changes in properties.  The change in mass, length, and width over time is shown in Figure~\ref{fig:AirSamples}.  The final results are summarized in Table~\ref{tab:samplechanges} for liquid exposures of nearly 200 days.

\begin{table}[tbp]
\def\pn{\phantom{-}}
   \centering
   \begin{tabular}{ll@{\hspace{3pt}}c@{\hspace{3pt}}ll@{\hspace{3pt}}c@{\hspace{3pt}}ll@{\hspace{3pt}}c@{\hspace{3pt}}l}
   \toprule
\emph{Property} 	& \multicolumn{9}{c}{\itshape Percent change in}\\
&\multicolumn{3}{c}{\itshape Deionized Water}	& \multicolumn{3}{c}{\itshape Linear Alkyl Benzene}	& \multicolumn{3}{c}{\itshape Mineral Oil} \\\midrule
Mass	     	& +$0.717$	& $\pm$	& $0.010$		& $-0.016$	& $\pm$	& $0.010 $	& $-0.047$	& $\pm$ & $0.010$ \\
Length   	& +$0.109$	& $\pm$	& $0.006$	& $\pn0.000$	& $\pm$	& $0.006$	& $-0.0162$	& $\pm$ & $0.006$\\
Width    	&  +$0.105$	& $\pm$	& $0.008$	& $-0.004$	& $\pm$	& $0.008$	& $-0.004$	& $\pm$ & $0.008$\\
Thickness	&  	     	& $<$	& $0.028$	&        	& $<$	& $0.028$	&        	& $<$	& $0.028$\\ \bottomrule
   \end{tabular}
   \caption{Summary of mass and dimensional changes to room-air moisture level acrylic samples after 200-day immersion in liquids.  The sample in DI water clearly shows an increase in mass, width, and length, while the properties of samples in MO and LAB remain relatively constant.}
   \label{tab:samplechanges}
\end{table}

These measurements clearly show significant absorption of DI water into the room-air moisture acrylic, with a final water-saturated mass of 1.6\% of total mass and an exponential time constant of roughly 400 days, in good agreement with~\cite{SNO}. More importantly, these tests show little change in acrylic sample mass after long-term exposure to Daya Bay detector liquids, whether from absorption of those liquids by the acrylic, or from desorption of water from the acrylic into the detector liquids.  This result supports the conclusion that ADs not stored for prolonged periods in a dry environment should experience negligible changes in target mass from AV-liquid interactions.  To be more precise, the change in target mass from nearly one year of exposure to detector liquids will be less than 0.016\% of the IAV's mass, which, at an IAV mass of roughly 1 ton, corresponds to 0.16~kg, or 0.0008\% of the 20-ton target mass.  This is clearly a negligible change in total target mass for these ADs from this source.

The samples soaked in LAB and MO showed no change in their transmittance spectrum within the measurement uncertainty of the spectrometer. Absorption of LAB and mineral oil into acrylic samples thus does not affect the transmittance of the acrylic at the percent level.

\subsubsection{Water-Saturated Acrylic}

In addition to studying acrylic at room-air moisture levels, interaction of water-saturated acrylic with detector liquids was investigated by placing two of the water-saturated samples from the previous test into the LAB test beaker and one water-saturated sample into the MO test beaker.  As with the previous test, the mass and dimensions of all of the samples were periodically measured.  Monitoring and measurement of the other samples already residing in the MO and LAB were continued during this period.

\begin{figure}[htb]
\subfloat[Change in mass of water-saturated samples submerged in MO.]{\label{subfig:SoakedMO}%
  \includegraphics[trim=0cm 0cm 0cm 1cm, clip=true, width=0.48\textwidth]{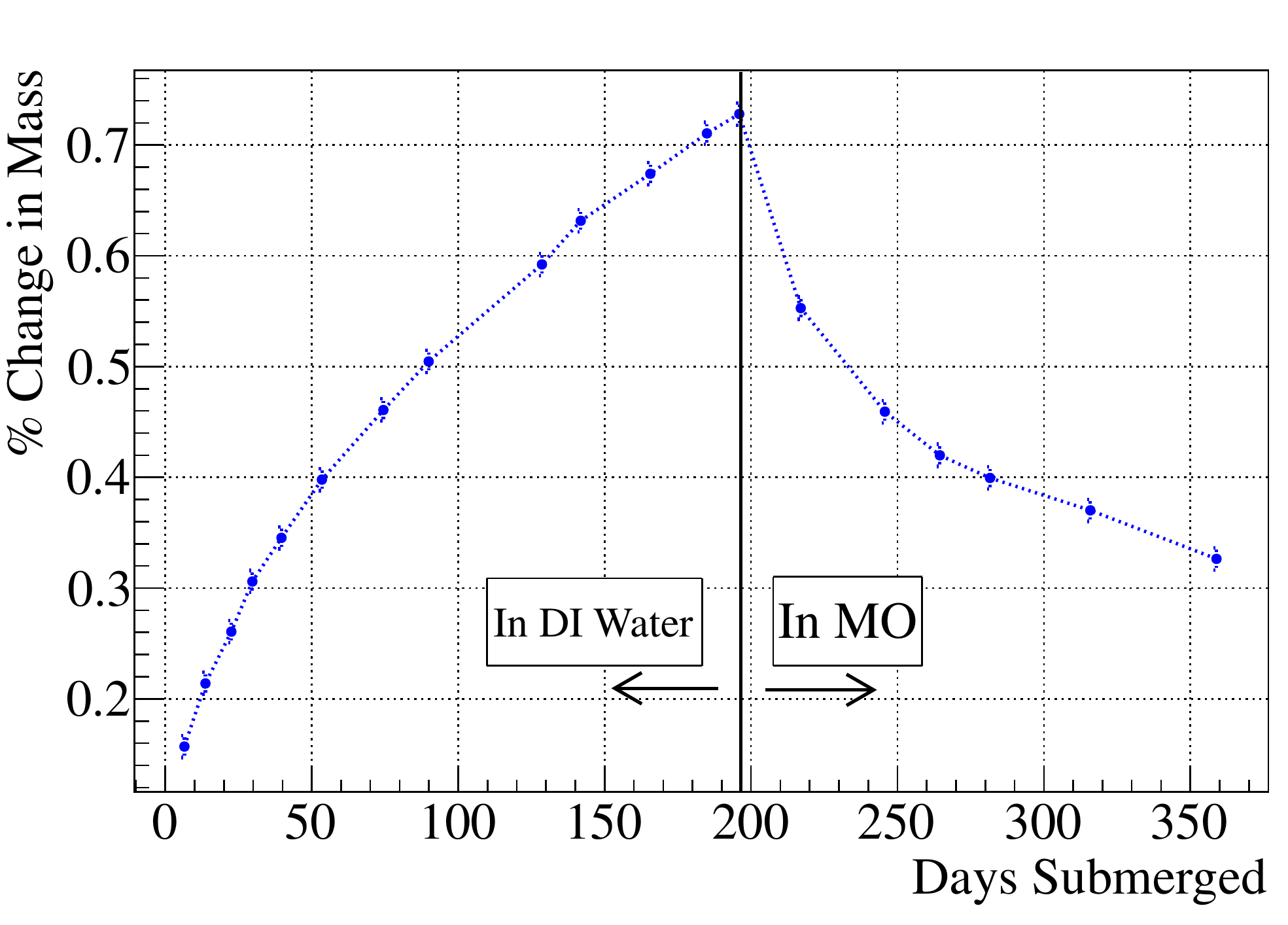}%
}\hfill
\subfloat[Change in mass of water-saturated samples submerged in LAB.]{\label{subfig:SoakedLAB}
  \includegraphics[width=0.48\textwidth]{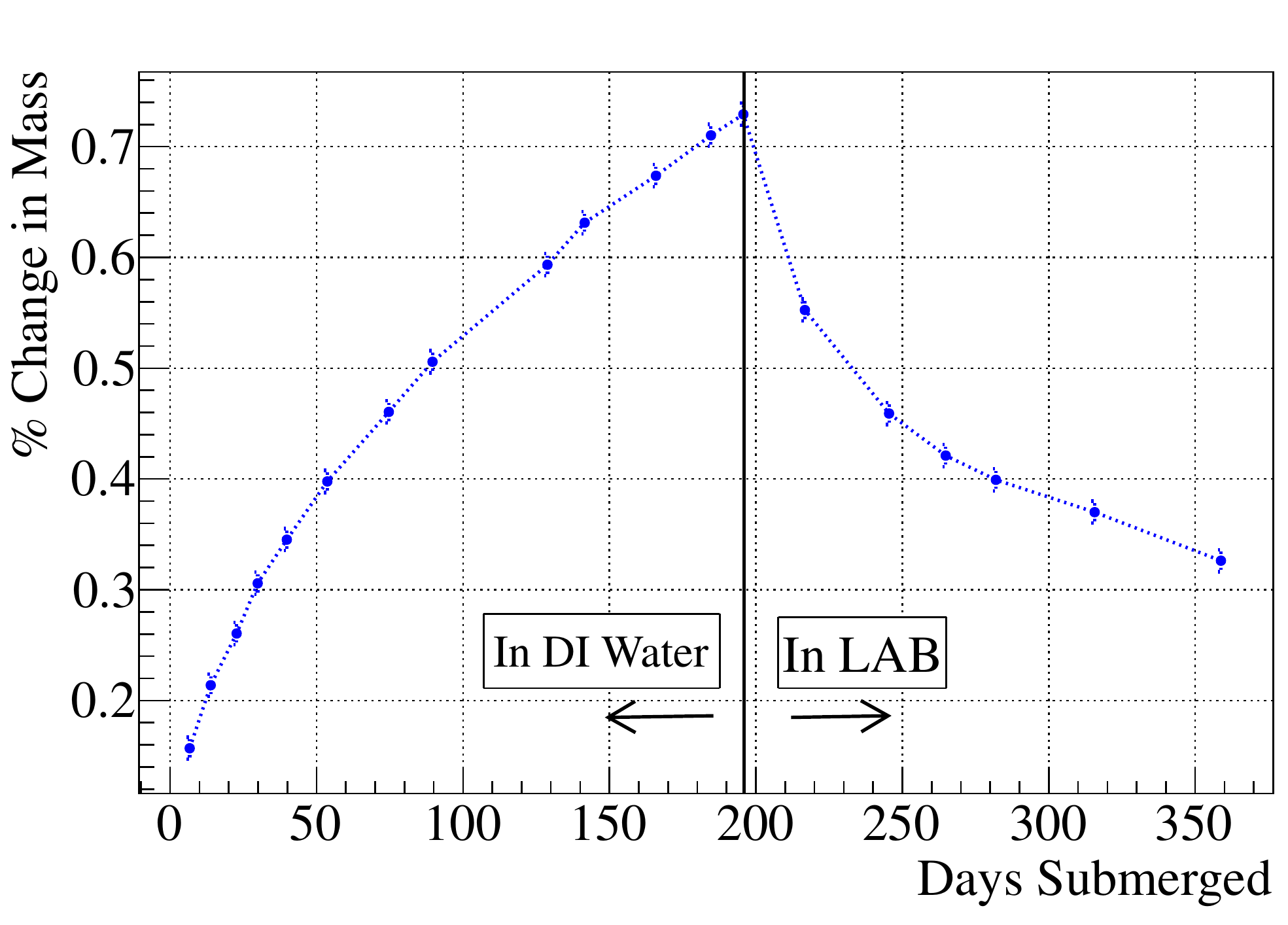}%
}%
\caption{Change in mass of room-air moisture level samples exposed first to a DI water environment and then to a MO or LAB environment. The point of switch-over is shown by the vertical line.  Samples clearly absorbed DI water and then desorbed this water over time in the detector liquids.}
\label{fig:SoakedPlots}
\end{figure}

These initially water-saturated samples showed a decrease in their mass and in their dimensions with time after being submerged in MO and LAB, as shown in Figure~\ref{fig:SoakedPlots}.   This trend indicates that the DI water that had saturated these pieces was seeping out when they were placed into MO or LAB.

\begin{figure}\centering
        \includegraphics[height=2.5in]{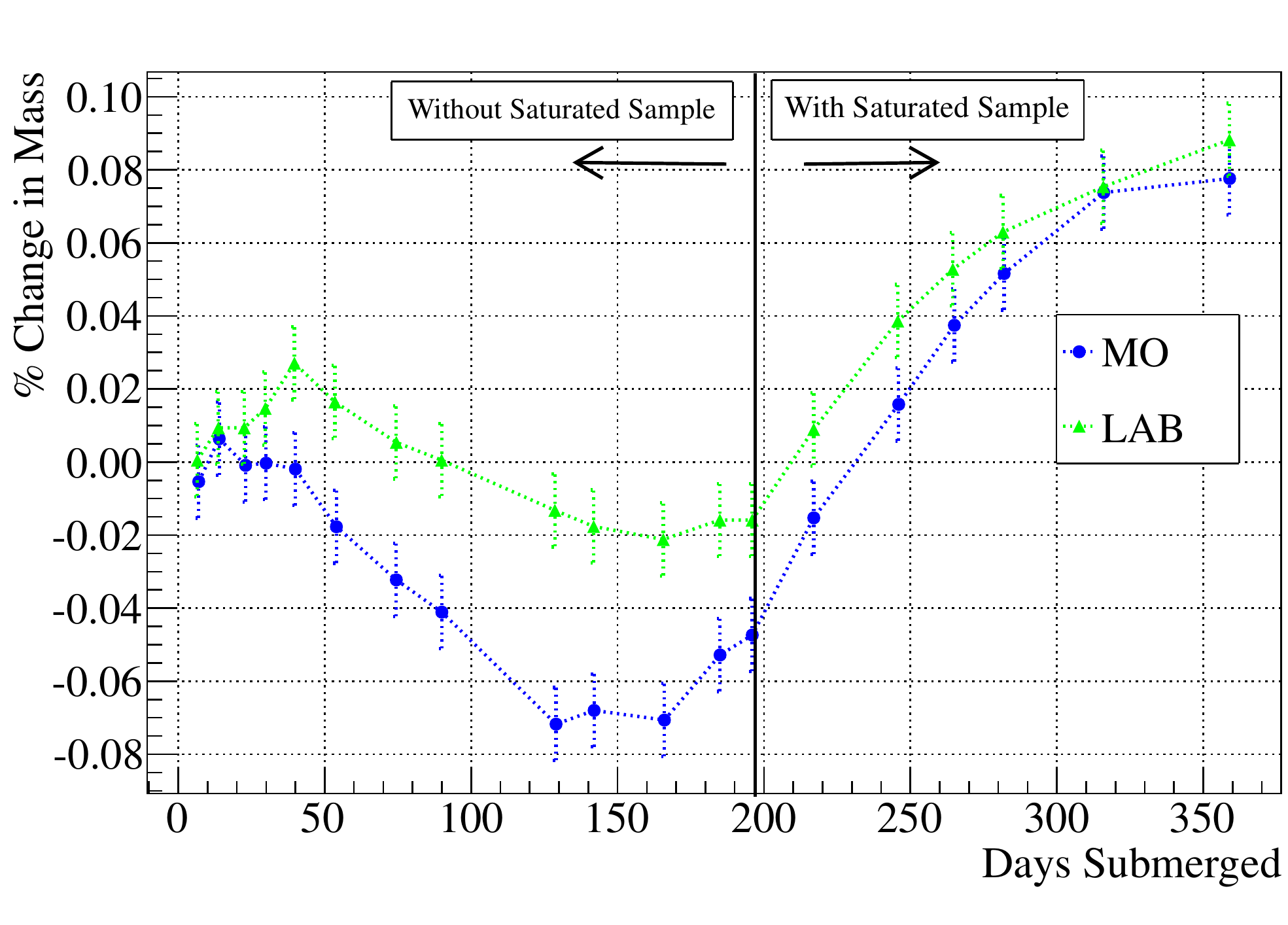}
        \caption{The change in mass of room-air moisture level samples submerged in MO and LAB.  The first 200 days are replicated from Figure~11a; after 200 days, water-soaked samples were transferred to the LAB and MO.  After this change, the original MO- and LAB-submerged acrylic samples show a slight increase in mass.  This increase could indicate transition of water from the water-soaked samples to the other samples, or could be a product of systematic environmental changes in the laboratory, such as a slow shift in room temperature or humidity.}
  \label{fig:ChangedLiquids}
\end{figure}

The final location of the water desorbed from these samples is uncertain.  Given the non-polar nature of the LAB and MO, it seems possible that the polar desorbed water molecules may have remained on the sample surface, and were removed during the preparation process for sample measurement.  Alternatively, the desorbed water may have dissolved into the detector liquids.  This hypothesis could be supported by the increase in mass of the air-moisture level samples that share the MO and LAB baths with the water-saturated samples during days 200--360, as shown in Figure~\ref{fig:ChangedLiquids}: the desorbed, dissolved water could subsequently be absorbed by these other comparatively low-moisture samples, raising these samples' measured masses.  However, this increase in mass is quite small, $<0.1\%$, and might instead have been caused by systematic changes in environment or in sample preparation over time.

As with the samples equilibrated to average room air humidity, no changes in transmittance were observed over time within the systematic uncertainty of the spectrometer.  In addition, the visual appearance of the detector liquids was unchanged over the duration of the test.  Thus, whatever the final destination of the excess water, the optical properties of the acrylic and detector liquids were not significantly altered, even in this extreme moisture level case.

\subsubsection{Dry Acrylic}
\label{subsec:Dry}

\begin{figure}\centering
        \includegraphics[height=2.5in]{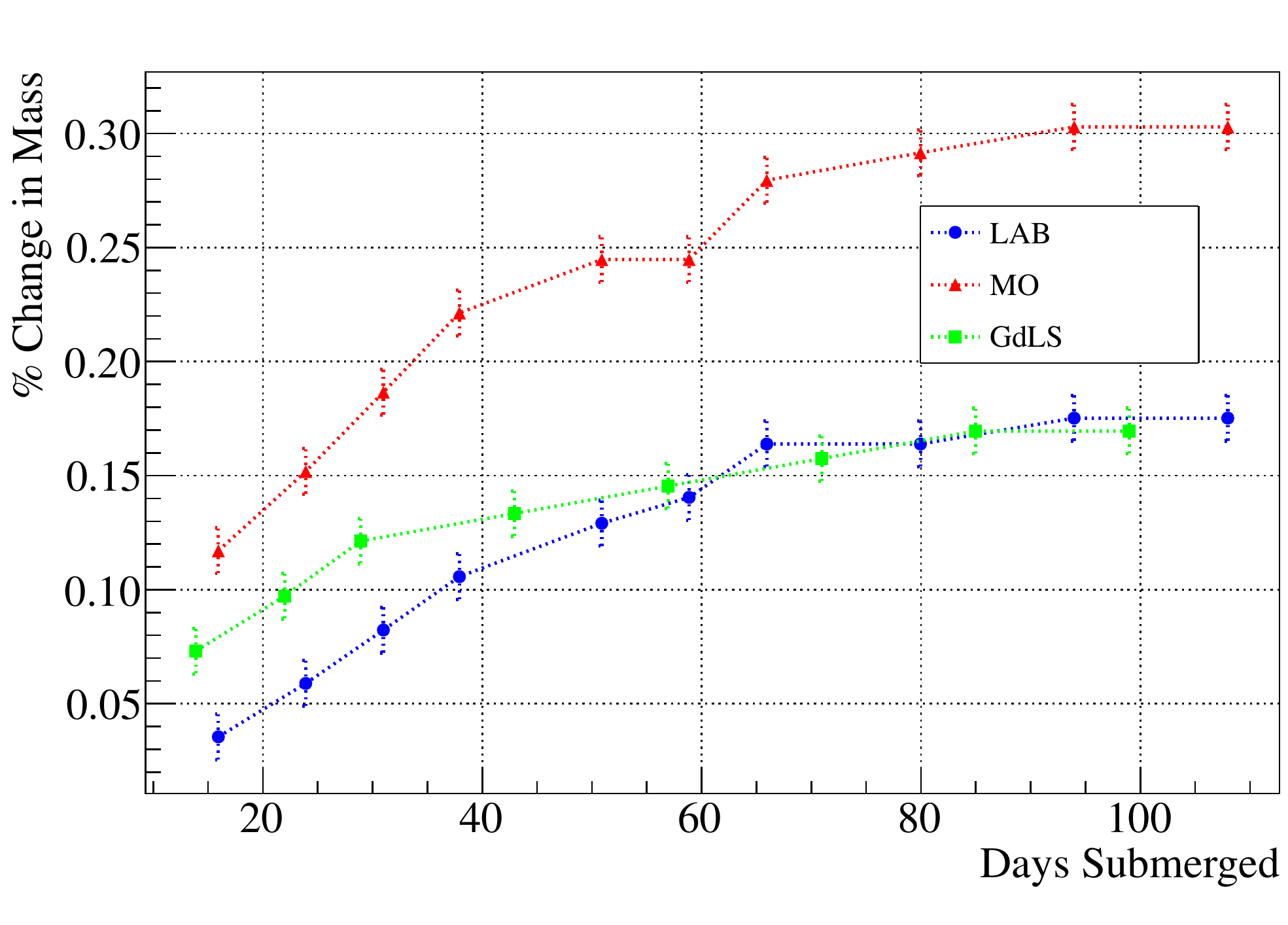}
        \caption{Percent change in the mass of acrylic samples dried via oven-baking before being submerged in the three different Daya Bay detector liquids.}
    \label{fig:DrySoak}
\end{figure}

To test the interaction of dry acrylic with detector liquids, the samples that had been baked and dried for a four-week period as described in Section~\ref{subsec:DryEnv} were placed in MO and LAB, as well as in Gd-LS obtained from the collaboration. The change in mass with time is given in Figure~\ref{fig:DrySoak}.  The sample in LAB increased its mass by $0.1752\pm 0.008\%$ over 110 days while the sample in mineral oil increased its mass by $0.3026\pm 0.008\%$ over the same period of time. The sample in Gd-LS increased its mass by $0.1689\pm 0.008\%$.  The relatively short time constant of this mass change indicates that any change in mass beyond 110 days of soaking will be minimal.  Despite this change in mass, the samples showed no change in transmittance over the duration of the test within the systematic uncertainty of the spectrometer.

After this test, it was unclear whether the increase in sample mass was the result of absorption of detector liquids, or of absorption of water during sample cleaning and measurement.  To clarify this, two other acrylic samples were dried for a week using the convection oven and then placed into LAB.  One sample was cleaned, measured, and replaced in LAB repeatedly over the course of the next two days, while the other was left undisturbed in the LAB.  Both samples were then removed from the liquid, cleaned and weighed, with both showing changes in mass of 0.01\% or less.  This indicates that prolonged exposure to detector liquids, rather than number of cleaning cycles, results in an increase in sample mass.

This test demonstrates that dry acrylic samples submerged in detector liquids absorb a small but non-zero amount of each type of detector liquid.  The target mass of the first two ADs, which contain IAVs with dry acrylic from nitrogen boil-off storage, will thus be slightly impacted by acrylic absorption of Gd-LS.  If the dry IAV only absorbs Gd-LS, the 0.17\% change in the IAV mass from Gd-LS absorption corresponds to a 0.0085\% change in the target mass of these AD.  This is a negligible contribution to the total target mass uncertainty of 0.03\%, although larger than the 0.0008\% variation experienced by the ADs with IAVs of normal moisture level.

It is unclear from these tests whether the dry acrylic absorbs LAB and MO molecules, or absorbs water known to be present in small amounts in the Gd-LS, LAB and MO.  Tests of production liquids done by Daya Bay collaborators report a water content of 23 ppm in LAB and roughly 150 ppm in Gd-LS~\cite{Minfang, Jun}.   In either case, the change in mass has clearly not impacted the acrylic optical properties or measured target mass in any meaningful way.

\section{Conclusions}
The acrylic target vessels for the Daya Bay experiment are critical components in the Daya Bay antineutrino detectors.  Their optical and mechanical stability is of paramount importance for the containment of the target liquids during the operation of the antineutrino detectors and in allowing the transmission of scintillation light from the antineutrino interactions.  This paper has described the effects of varied environmental conditions on the optical, mechanical, and material properties of the acrylic:

\begin{itemize}
\item{\textbf{Temperature Fluctuations During Shipping:} Typical temperature fluctuations from 10--40$^{\circ}$C experienced during shipping and simulated in the laboratory with acrylic samples in a programmable convection oven appear to have no impact on the mechanical or optical properties of the Daya Bay acrylic.}
\item{\textbf{Storage in a Low-Humidity Environment:} The dry nitrogen gas environment used to store some Daya Bay detector vessels after assembly was simulated by drying acrylic samples in a convection oven and in a nitrogen-flushed vacuum chamber.  Acrylic samples saturated with average room air humidity exhibited significant moisture loss of roughly 0.1\% of total mass over month-long timescales.  Such a loss of moisture should be expected in dry-stored acrylic vessels.  The resulting acrylic dryness has negligible impact on the optical and mechanical properties of the Daya Bay acrylic.}
\item{\textbf{Prolonged Exposure to Detector Liquids:} Acrylic samples with varying moisture content were submerged in various detector liquids to simulate the AV environment during detector operation.  All acrylic samples showed no variability in optical properties resulting from this exposure.  In addition, all samples exhibited minimal ability to absorb the detector liquids, LAB, MO, and Gd-LS, ranging from 0.016\% absorption by mass of LAB for samples at room air moisture to 0.3\% absorption of MO by dry samples.  AV absorption of the Gd-LS target liquid can thus alter the total AD target mass by a maximum of 0.0085\%, a small contribution to the detector's target mass uncertainty of 0.03\%.}
\end{itemize}

These tests present the Daya Bay acrylic as a pleasingly inert detector material.  It shows little change in optical or mechanical performance in a wide range of environments of varying temperature and moisture level.  In addition, the acrylic shows little affinity for absorption or alteration of the Daya Bay detector liquids.  These tests provide strong confidence in the Daya Bay acrylic vessels' long-term optical and mechanical stability and their ability to maintain the measured precision of Daya Bay's target masses.

\section*{Acknowledgements}

This work was done with support from the Department of Energy, Office of Science, High Energy Physics, and the University of Wisconsin. We thank Reynolds Polymer Technology for providing the acrylic samples studied in this paper and Tom Wise for support in setting up the experimental tests.  We are grateful to Paul Hinrichs, David Webber, and Tom Wise for careful reading of this manuscript, and to Jun Cao and Minfang Yeh for providing information about the Daya Bay detector liquids.

\end{document}